\documentclass[prl,a4paper,twocolumn,groupedaddress,english]{revtex4-1}
\usepackage{babel}
\usepackage{xcolor}
\usepackage{graphicx}
\usepackage{amsmath}
\usepackage{amssymb}
\usepackage{hhline}

\newcommand{\bc}{\begin{center}}
\newcommand{\ec}{\end{center}}
\newcommand{\be}{\begin{equation}}
\newcommand{\ee}{\end{equation}}
\newcommand{\bea}{\begin{eqnarray}}
\newcommand{\eea}{\end{eqnarray}}

\definecolor{darkblue}{rgb}{0.1,0.2,0.6}
\definecolor{darkred}{rgb}{0.8,0.1,0.2}
\usepackage[colorlinks,citecolor=darkblue,linkcolor=darkred,urlcolor=darkblue]
{hyperref} 
\usepackage[all]{hypcap} 

\makeatletter
\adddialect\l@English\l@english
\makeatother

\definecolor{commentcolor}{rgb}{0.1,0.2,0.6}
\definecolor{commentcolor2}{rgb}{1,0,0}
\definecolor{commentcolorF}{rgb}{0,0,1}
\definecolor{commentcolorD}{rgb}{1,0.1,.1}
\definecolor{todocolor}{rgb}{0.8,0.1,0.2}

\begin{document}
\title{Critical Properties of the Superfluid -- Bose Glass Transition in Two Dimensions}
\author{Juan Pablo \'Alvarez Z\'u\~niga}
\author{David J. Luitz}
\affiliation{Laboratoire de Physique Th\'eorique, IRSAMC, Universit\'e de 
Toulouse,
{CNRS, 31062 Toulouse, France}}
\author{Gabriel Lemari\'e}
\author{Nicolas Laflorencie}
\affiliation{Laboratoire de Physique Th\'eorique, IRSAMC, Universit\'e de 
Toulouse,
{CNRS, 31062 Toulouse, France}}
\date{December 15, 2014}

\begin{abstract}
We investigate the superfluid (SF) to Bose glass (BG) quantum phase transition using extensive quantum Monte Carlo simulations of two-dimensional 
hard-core bosons in a random box potential. $T=0$ critical properties are studied 
by thorough finite-size scaling of condensate and SF densities, 
both vanishing at the same critical disorder $W_c=4.80(5)$.
Our results give the following estimates 
for the critical exponents: $z=1.85(15)$, $\nu=1.20(12)$, $\eta=-0.40(15)$.
Furthermore, the probability distribution of the SF response $P(\ln\rho_{\rm 
sf})$ displays striking differences across the transition: while it narrows with increasing system sizes $L$ in the SF phase, it broadens  
in the BG regime, indicating an absence of self-averaging, and at the critical point $P(\ln\rho_{\rm 
sf}+z \ln L)$ is scale invariant. Finally, high-precision
measurements of the local density rule out a percolation picture for the SF-BG transition.
\end{abstract}
\maketitle
\noindent{\it{Introduction---}}
The interplay between disorder and interactions in condensed matter systems, 
while intensively studied during the last decades, remains today puzzling in 
many respects for both experimental and theoretical investigations
\cite{giamarchi_foreword_2013}. First raised by experiments in the late 1980s 
on superfluid $^4$He in porous media~
\cite{finotello_sharp_1988,reppy_superfluid_1992}, the theoretical question of 
interacting bosons in the presence of disorder has been addressed at the same 
time by several pioneer works~
\cite{ma_localized_1985,ma_strongly_1986,giamarchi_localization_1987,
fisher_onset_1988,fisher_boson_1989}. It was then rapidly understood that for two 
dimensional (2D) bosons with repulsive interaction, superfluidity is robust to 
weak disorder. 

A breakthrough came with the 
thorough study of the critical properties of the quantum ($T=0$) phase 
transition between superfluid (SF) and localized Bose-glass (BG) regimes by 
Fisher {\it{et al.}}~\cite{fisher_boson_1989}. In particular, a generalization of the 
Josephson scaling relations~\cite{josephson_relation_1966} was given, thus 
predicting new critical exponents (see first line of Tab.~\ref{tab:1}). Following this 
work a great endeavour has been made, using exact numerical techniques such as 
quantum Monte Carlo (QMC)
\cite{krauth_superfluid-insulator_1991,scalettar_localization_1991,
sorensen_universal_1992,makivic_disordered_1993,weichman_comment_1995,
zhang_quantum_1995,kisker_bose-glass_1997,
alet_cluster_2003,
prokofev_superfluid-insulator_2004,priyadarshee_quantum_2006,
hitchcock_bose-glass_2006,gurarie_phase_2009,
carrasquilla_characterization_2010,lin_superfluid-insulator_2011,
soyler_phase_2011,meier_quantum_2012,
yao_critical_2014} or the density matrix renormalization 
group (DMRG)
\cite{pai_one-dimensional_1996,rapsch_density_1999,roux_quasiperiodic_2008},  in order to explore in detail the phase diagram of the disordered 
Bose-Hubbard model.
Nevertheless, a general consensus regarding the precise values of the critical exponents at the SF-BG 
transition is still lacking, despite huge analytical~\cite{giamarchi_localization_1987,fisher_boson_1989,
singh_real-space_1992,herbut_dual_1997,weichman_critical_2007,
altman_superfluid-insulator_2010,ristivojevic_phase_2012,
iyer_mott_2012,alvarez_zuniga_bose-glass_2013} and numerical~\cite{sorensen_universal_1992, makivic_disordered_1993,weichman_comment_1995,
zhang_quantum_1995,
kisker_bose-glass_1997,alet_cluster_2003,
prokofev_superfluid-insulator_2004,priyadarshee_quantum_2006,
hitchcock_bose-glass_2006,soyler_phase_2011,
meier_quantum_2012,yao_critical_2014} 
efforts.

At the same time a wealth of new experiments have been developed, using 
different techniques and setups:
(i) ultracold bosonic atoms in a random 
potential~\cite{white_strongly_2009,deissler_delocalization_2010,
krinner_superfluidity_2013,
derrico_observation_2014};
(ii) strongly disordered superconducting films where preformed 
Cooper pairs can localize~\cite{sacepe_localization_2011,
lemarie_universal_2013, PhysRevLett.109.107003,mondal_enhancement_2013};
(iii) impurity doped quantum magnets at 
high field~\cite{hong_evidence_2010,
huvonen_field-induced_2012,yu_bose_2012,vojta_excitation_2013,
zheludev_dirty-boson_2013}. 
They all have shed a new light on the problem of boson 
localization but raised important theoretical questions, regarding {\it{e.g.}} the precise 
nature of the critical point
\cite{weichman_critical_2007,
altman_superfluid-insulator_2010,ristivojevic_phase_2012,iyer_mott_2012}, 
the inhomogeneous character of the SF and BG phases
\cite{feigelman_superconductor-insulator_2010,sacepe_localization_2011,lemarie_universal_2013,krinner_direct_2013}.

In this Letter, we address two important issues of the Bose glass problem using 
the most advanced
available exact numerical technique, namely the stochastic series expansion (SSE)
QMC method. 
The quantum critical behavior at
the onset of boson localization and the delicate estimate of the critical exponents 
are first
discussed. Then the inhomogeneous nature of the SF and BG phases is addressed through 
the study of the
probability distribution of the SF response which shows strikingly different properties when increasing lattice sizes. Shrinking in the SF phase, it clearly broadens in the BG regime, thus indicating the absence of self-averaging~\cite{hegg_breakdown_2013}. We also 
demonstrate that all sites remain compressible, ruling out a percolation 
picture. Our conclusions are supported by careful ground-state (GS) simulations 
through the so-called $\beta$-doubling scheme, disorder averaging over a very
large number of
realizations, 
detailed error bar evaluation, and systematic finite-size scaling analysis.

\noindent{\it Model and Quantum Monte Carlo approach---} 
We consider hard-core bosons at half-filling on a two-dimensional 
square lattice, described by
\be
{\cal{H}}=-t\sum_{\langle ij\rangle} \left(b^{\dagger}_{i}b^{\vphantom{\dagger}}_{j}
+b^{\dagger}_{j}b^{\vphantom{\dagger}}_{i}\right)-\sum_i \mu_i b^{\dagger}_{i}
b^{\vphantom{\dagger}}_{i},
\label{eq:1}
\ee
where hopping between nearest neighbours is fixed to $t=1/2$, and the random 
chemical potential $\mu_i$ is drawn from a uniform distribution $[-W,W]$, 
{\it{i.e.}} half-filling is statistically achieved, on average~\cite{weichman_particle-hole_2008}. This model, also relevant to describe many aspects of strongly disordered 
superconductors~\cite{ma_localized_1985,ma_strongly_1986,feigelman_superconductor-insulator_2010,sacepe_localization_2011,seibold_superfluid_2012,lemarie_universal_2013}, exhibits a quantum ($T=0$) phase 
transition between a Bose condensed SF and a localized BG regime at 
sufficiently strong disorder~\cite{makivic_disordered_1993,zhang_quantum_1995,priyadarshee_quantum_2006}.

    \begin{figure}[t] \centering \includegraphics[width=\columnwidth,clip]{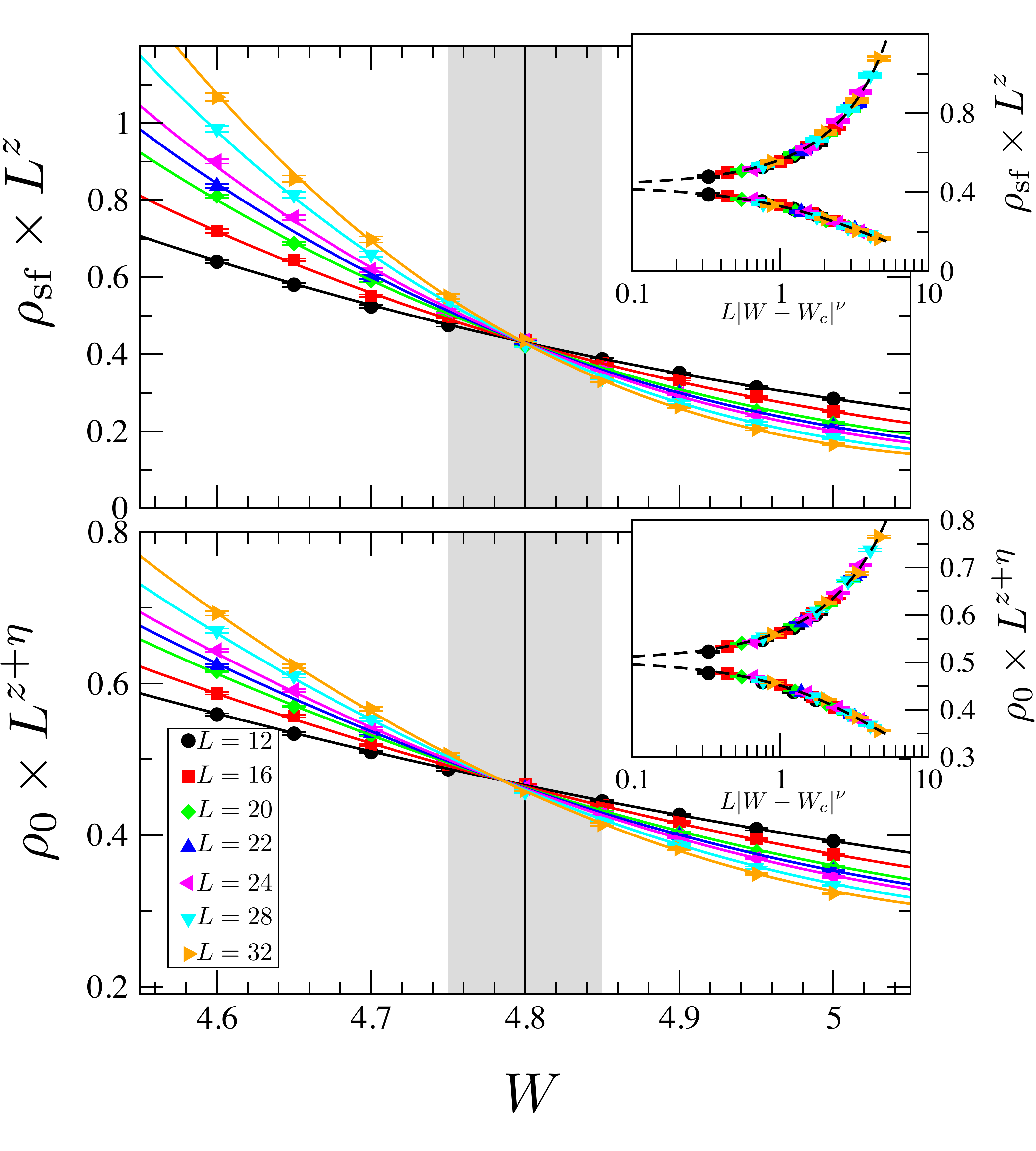}
        \caption{Scaling analysis of the SF $\rho_\text{sf}$ (top) and 
BEC $\rho_0$ (bottom) densities. Solid lines show best fits to the universal scaling functions 
Eqs.~\eqref{eq:gsf} and \eqref{eq:g0} for the full data set with
$z\simeq 1.85$, $W_c^{\rm sf}\simeq 4.8$, $W_c^{\rm 0}\simeq 4.79$, $z+\eta\simeq 1.42$, 
$\nu_{\rm sf}\simeq 1.1$, $\nu_{0}\simeq 1.2$, and $\mathcal G_{\text{sf}|
0}$ $3^{\rm rd}$
        order polynomials. 
The distance from the critical point $W_c=4.80(5)$ 
(grey area), when rescaled by $L^{-1/\nu}$ with $\nu=1.2$, yields a perfect 
data collapse (insets).}
\label{fig:2} \end{figure}

The intrinsic 
difficulties to simulate with QMC methods the low temperature properties of such a strongly disordered quantum system are twofold: 
(i) accessing ground-state (GS) 
properties means very long equilibration and simulation times; (ii) 
statistical uncertainties of the measured physical 
observables originate from both MC sampling with $N_{\rm mc}$ steps and random sample to sample fluctuations with 
${\cal N}_{\rm s}$ samples. Therefore, the simulation time grows very fast as $L^2\times \beta\times N_{\rm mc}\times {\cal{N}}
_{\rm s}$  which limits the largest system size $L$ reachable. 
The strategy we adopt to tackle this problem, using the SSE algorithm~\cite{syljuasen_quantum_2002}, is as follows (simulation details are discussed in the supplementary 
material \cite{sm}).
First we use the $\beta$-doubling scheme to 
speed up equilibration towards very low temperature
\cite{sandvik_classical_2002,laflorencie_random-exchange_2006,sm}, after what we perform for each sample a number of
measurement steps $N_{\rm mc}^s$ (sample dependent) large enough that the SF density is efficiently measured~\cite{sm}.
This procedure is then repeated for a very large 
number of disorder realizations ${\cal N}_{\rm s}={\cal{O}}(10^4)$. We have noticed that GS convergence 
is in practice extremely hard to achieve rigorously for 
all samples, as some samples may exhibit finite-size gaps
smaller than the infrared cutoff of the $\beta$-doubling 
expansion, which is fixed on average. Nevertheless, we have checked that intrinsic MC errors induced by such a slow GS convergence remain 
smaller than statistical errors. The results, at $\beta t=2^h$ with
$h=7$ for $L=12$ up to $h=9$ for the largest sizes, can therefore be safely interpreted as 
$T=0$ ones~\cite{sm}.

\begin{table}[b]
    \begin{tabular}{llll|r}
$z$&$\nu$&$\phantom{-}$$\eta$&$W_c$&Reference\\
\hline
$2$&$\ge 1$&$\phantom{-}$$\le 0$&&Fisher  {\it{et al.}}~\cite{fisher_boson_1989} \\ 
\hline
$0.5(1)$&$2.2(2)$&$\phantom{-}$n.a.&$2.5$& Makivi\'c {\it{et al.}}
\cite{makivic_disordered_1993}\\
$2.0(4)$&$0.90(13)$&$\phantom{-}$n.a.&$4.95(20)$&Zhang {\it{et al.}}
\cite{zhang_quantum_1995}\\
$1.40(2)$&$1.10(4)$&$-0.22(6)$&$4.42(2)$&Priyadarshee {\it{et al.}}
\cite{priyadarshee_quantum_2006}\\
\hline
$1.85(15)$&$1.20(12)$&$-0.40(15)$&$4.80(5)$& This work\\
    \hline
    \hline
\end{tabular}
\caption{\label{tab:1}Various estimates of critical exponents and disorder strength $W_c$ for the 2D SF--BG 
transition of model Eq.~\eqref{eq:1}.}
\end{table}

\noindent{\it{Finite-size scaling---}} 
Motivated by the fact that previous works disagree on the values of the critical parameters 
(see~\cite{makivic_disordered_1993,zhang_quantum_1995,priyadarshee_quantum_2006} and Tab.~\ref{tab:1}), we now discuss our determination of these parameters by the finite-size scaling approach for disorder averaged QMC 
estimates of the SF and Bose condensed densities. 

The ordered regime is
characterized by a finite SF density
$\rho_{\rm sf}$, efficiently estimated using the winding number 
fluctuations in the QMC algorithm~\cite{Ceperley87}. In the vicinity of the 2D 
quantum critical point, the finite-size scaling of the SF density is
\be
\rho_{\rm sf}(L)=L^{-z} \; \mathcal G_{\rm sf}[L^{1/\nu}(W-W_c)],
\label{eq:gsf}
\ee
where $z$ is the dynamical critical exponent, $\nu$ the correlation length 
exponent, $W_c$ the critical disorder, and $\mathcal G_{\rm sf}$ a universal function. 

Beyond the SF response, 
one can also probe Bose-Einstein condensation (BEC), occurring in 2D 
at $T=0$ where U(1) symmetry can be broken. The BEC density $\rho_0=
\sum_{ij}G_{ij}/N^2$, 
obtained from the equal time Green's function~\cite{Dorneich01} $G_{ij}=\langle 
b^{\dagger}_{i}b^{\vphantom{\dagger}}_j\rangle$, plays the role of the order parameter, with a critical scaling
\be
\rho_{0}(L)=L^{-z-\eta} \; \mathcal G_{0}[L^{1/\nu}(W-W_c)].
\label{eq:g0}
\ee
Our QMC data are very nicely described by the above scaling forms, as shown in Fig.~\ref{fig:2} for both
SF and BEC densities. Strikingly,
BEC and SF densities vanish at the same disorder strength $W_c=4.80(5)$. The values of the critical exponents are given in Table~\ref{tab:1}. 
This determination results from fits of our data set by Taylor expanding the scaling functions $\mathcal G_{\rm sf}$ and $\mathcal G_{0}$ around $W_c$ up to an order large enough that the goodness of fit is acceptable (3rd order in Fig.~\ref{fig:2}, see~\cite{sm}). 
We have performed a careful error analysis using the bootstrap
approach in order to estimate statistical errors of the fit parameters, as well as potential
systematic errors by fitting over various ranges of disorder strengths and sizes~\cite{sm}. This results in
conservative uncertainties for the estimates of the critical parameters, as visible in Tab.~\ref{tab:1}.

We observe a good agreement with the predicted bounds from Fisher {\it{et al.}}~\cite{fisher_boson_1989} for $\nu=1.20(12)\ge 1$ and $\eta=-0.40(15)\le 0$. Regarding the more debated question of the dynamical exponent~\cite{weichman_critical_2007}, while still compatible with $z=2$ within error bars our best estimate gives a smaller number $z=1.85(15)$, in agreement with a recent careful estimate for quantum rotors~\cite{meier_quantum_2012}. Comparing with other studies in Tab.~\ref{tab:1}, our results, obtained with much larger system sizes, agree within error bars with Ref.~\cite{zhang_quantum_1995}, whereas results in 
Refs.~\cite{makivic_disordered_1993,priyadarshee_quantum_2006} are probably biased due to finite temperature effects and too small disorder averaging.

    \begin{figure*}[t] \centering \includegraphics[width=2\columnwidth,clip]{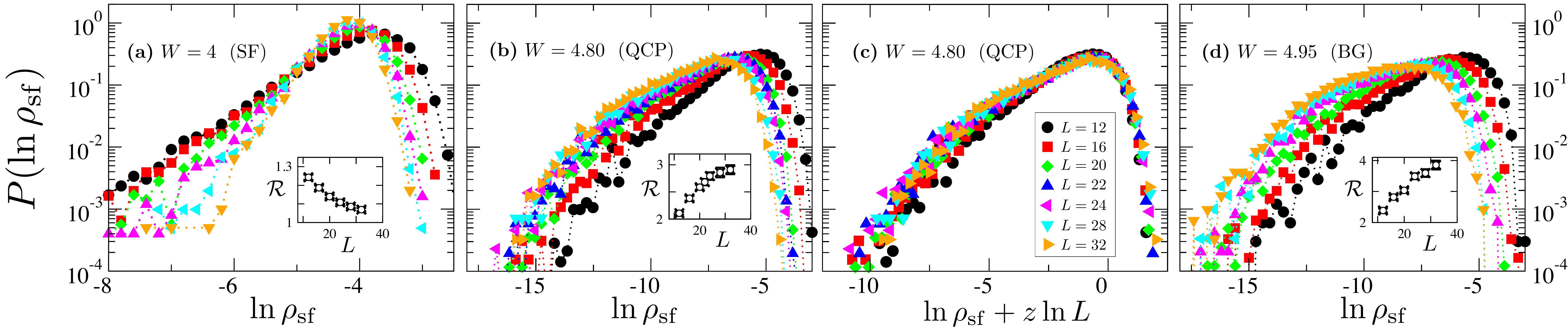}
       \caption{Histogram of QMC estimates for $\ln\rho_{\rm sf}$ performed over 
       ${\cal{N}}_{\rm s}\sim 10^4$ disordered samples for each size $L$. (a) In the SF regime $W=4$, distributions get narrower with increasing $L$ whereas in the BG phase (d) for $W=4.95$ they broaden. At criticality $W_c=4.8$ (b-c) the broadening stops above $L=20$ and $P(\ln\rho_{\rm sf}+z\ln L)$ displays a good collapse using $z=1.85$. The insets show the ratio ${\cal R}={\rho_{\rm sf}^{\rm avg}}/{\rho_{\rm sf}^{\rm typ}}$ {\it{vs.}} system size $L$.}
\label{fig:3} \end{figure*}

\noindent{\it{Distributions and absence of self-averaging in the BG---}} In order to go beyond the analysis of the critical properties based on disorder averaged observables, we now turn to the much less studied issue of distributions. 
The question of a possible broadening of the responses, linked to the issue of self-averaging, has not been studied for 2D bosons, although it may be crucial as discussed for disordered Ising models~\cite{fisher_critical_1995,young_numerical_1996,wiseman_finite-size_1998} and strongly disordered superconductors \cite{feigelman_superconductor-insulator_2010,sacepe_localization_2011,lemarie_universal_2013}.
Here we focus on the probability distribution $P(\ln \rho_{\rm sf})$, obtained by building histograms
of QMC estimates for $\ln \rho_{\rm sf}$ 
over $\cal{N}_{\rm s}$ independent samples, with 
${\cal N}_{\rm s}\approx 2\times 10^4$ for $L\le 22$ and ${\cal N}_{\rm s}\approx 10^4$ for $L\ge 24$, 
shown in Fig.~\ref{fig:3} for three values of the disorder strength. 

In the SF regime (panel (a) $W=4<W_c$) 
the distribution narrows upon increasing the size $L$, thus demonstrating that the SF response is self-averaging in the ordered phase. 
Conversely, as visible in panel (d) for the BG regime at $W=4.95>W_c$, $P(\ln\rho_{\rm sf})$ 
broadens when $L$ increases, and moves towards large negative values, as expected in the 
thermodynamic limit where the SF stiffness vanishes. We therefore expect a difference between
average and typical SF densities in the BG: as shown in the insets of Fig.~\ref{fig:3}, the ratio
${\cal R}={\rho_{\rm sf}^{\rm avg}}/{\rho_{\rm sf}^{\rm typ}}$ clearly increases with $L$ in the BG regime
(d) whereas it goes to 1 in the SF phase (a). 
At the critical point $W_c=4.8$ (panels (b-c) of
Fig.~\ref{fig:3}), the histograms first broaden for small sizes and then, above $L=20$ the curves
appear self-similar, simply shifted relative to each other.  This absence of broadening at large
scales is also visible in the inset (b) where the ratio ${\cal R}$ tends to saturate to a constant value. 
The shift of the distributions can be
corrected for by adding $z\ln L$ to $\ln \rho_{\mathrm{sf}}$ using our best estimate $z=1.85$. Indeed, as shown in Fig.~\ref{fig:3} (c) $P[\ln(L^z\rho_{\rm sf})]$ yields a collapse onto a scale invariant distribution, particularly good above $L=20$.

    \begin{figure}[b] \centering \includegraphics[width=\columnwidth,clip]{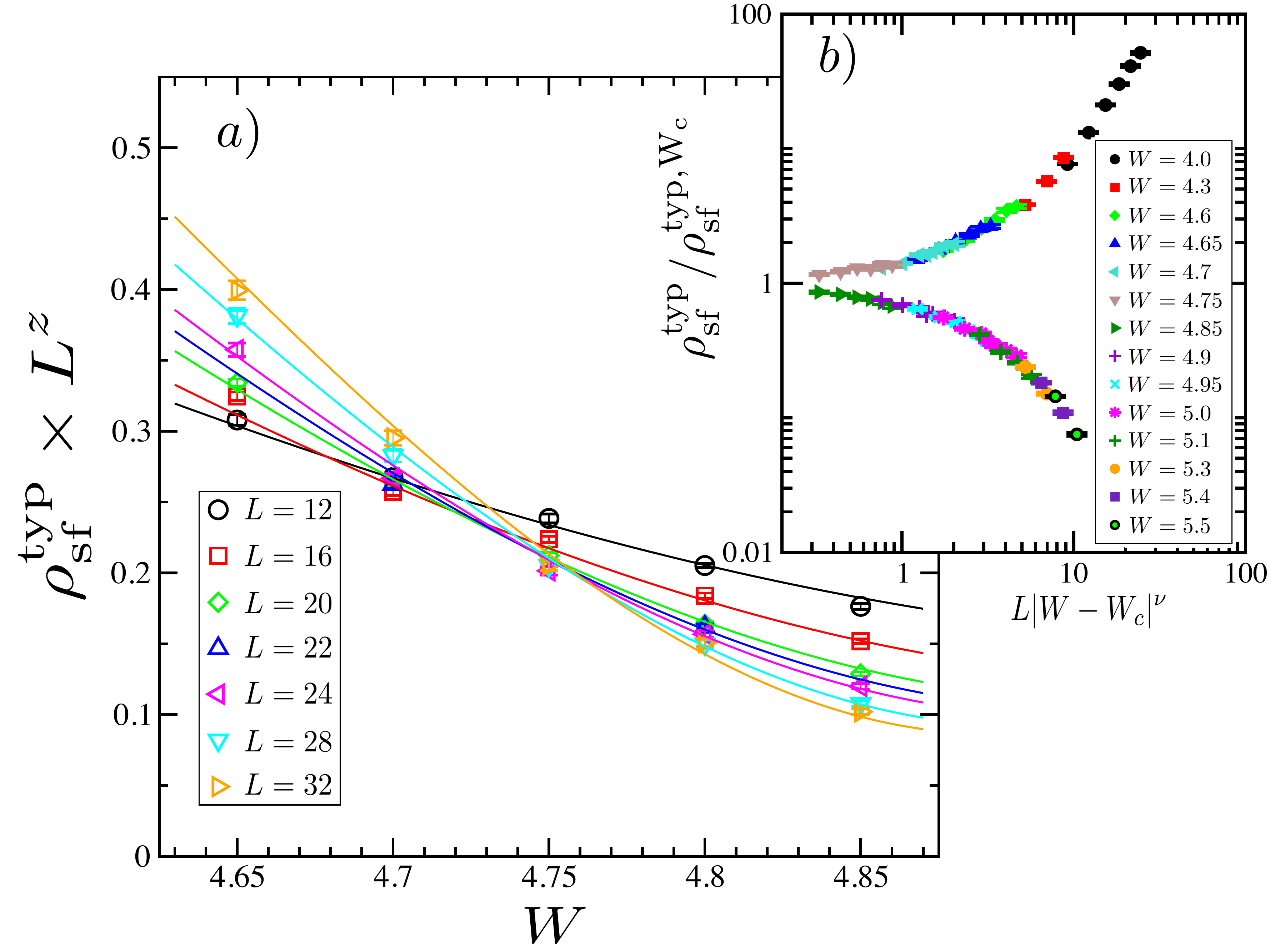}
       \caption{{Typical SF density (a) plotted as $\rho_{\rm sf}^{\rm typ}\times L^z$ {\it{vs.}} $W$ where the crossing at $W_c=4.8$ has a visible drift, captured by ${\cal{G}}_{\rm sf}^{\rm typ}[L^{1/\nu}(W-W_c)]+cL^{-y}$ with fixed $\nu=1.2$ and $z=1.85$, and an estimated irrelevant exponent $y=0.97(4)$. In panel (b), $\rho_{\rm sf}^{\rm typ}/\rho_{\rm sf}^{{\rm typ}, W_c}$ plotted against $L|W-W_c|^\nu$ exhibits an almost perfect collapse of the data for $4\le W\le 5.5$ and $12\le L\le 32$ with no additional parameters.}}
\label{fig:4} \end{figure}

The fact that all distributions at $W_c$ are identical up to a shift suggests that, while typical and
average SF densities scale differently in the BG regime, their critical scalings are described by
the same exponents. Indeed, the typical SF density, defined as $\rho_{\rm sf}^{\rm
typ}=\exp({\overline{\ln\rho_{\rm sf}}})$ (where ${\overline{(\cdots)}}$ stands for disorder
averaging), can be analyzed using a scaling hypothesis similar to the average Eq.~\eqref{eq:gsf}, but including additional irrelevant corrections~\cite{rodriguez_multifractal_2011}
\be
\rho_{\rm sf}^{\rm typ}(L)=L^{-z}\left(\mathcal G_{\rm sf}^{\rm typ}[L^{1/\nu}(W-W_c)]+cL^{-y}\right).
\label{eq:gsftyp}
\ee
Because of the presence of irrelevant corrections, a fit of our data set by Eq.~\eqref{eq:gsftyp} with a polynomial $\mathcal G_{\rm sf}$ is unstable unless we fix the critical parameters $W_c$, $z$ and $\nu$ to our best estimates (Tab.~\ref{tab:1}). The crossing of $\rho_{\rm sf}^{\rm typ}\times L^z$ vs $W$
plotted in Fig.~\ref{fig:4} (a) displays a non-negligible drift, well captured by irrelevant corrections in Eq.~\eqref{eq:gsftyp} with $y=0.97(4)$. A nice way to achieve a scaling plot for the typical SF density is then to divide $\rho_{\rm sf}^{\rm typ}$ by its value at $W_c$, this in order to cancel out the irrelevant corrections $\sim L^{-y}$. Next, a rescaling of the length $L$ by the correlation length $\xi=|W-W_c|^{-\nu}$ with $\nu=1.2$ and $W_c=4.8$, gives an almost perfect collapse, without any additional adjustable parameters, as shown in Fig.~\ref{fig:4} (b) for $4.0\le W\le 5.5$ and all available system sizes $12\le L\le 32$. This demonstrates that the quantum critical behaviours of average and typical SF densities are similar, in particular their critical exponents $z_{\rm avg}=z_{\rm typ}=1.85(15)$ and $\nu_{\rm typ}=\nu_{\rm avg}=1.20(12)$.

    \begin{figure}[b] \centering \includegraphics[width=.95\columnwidth,clip]{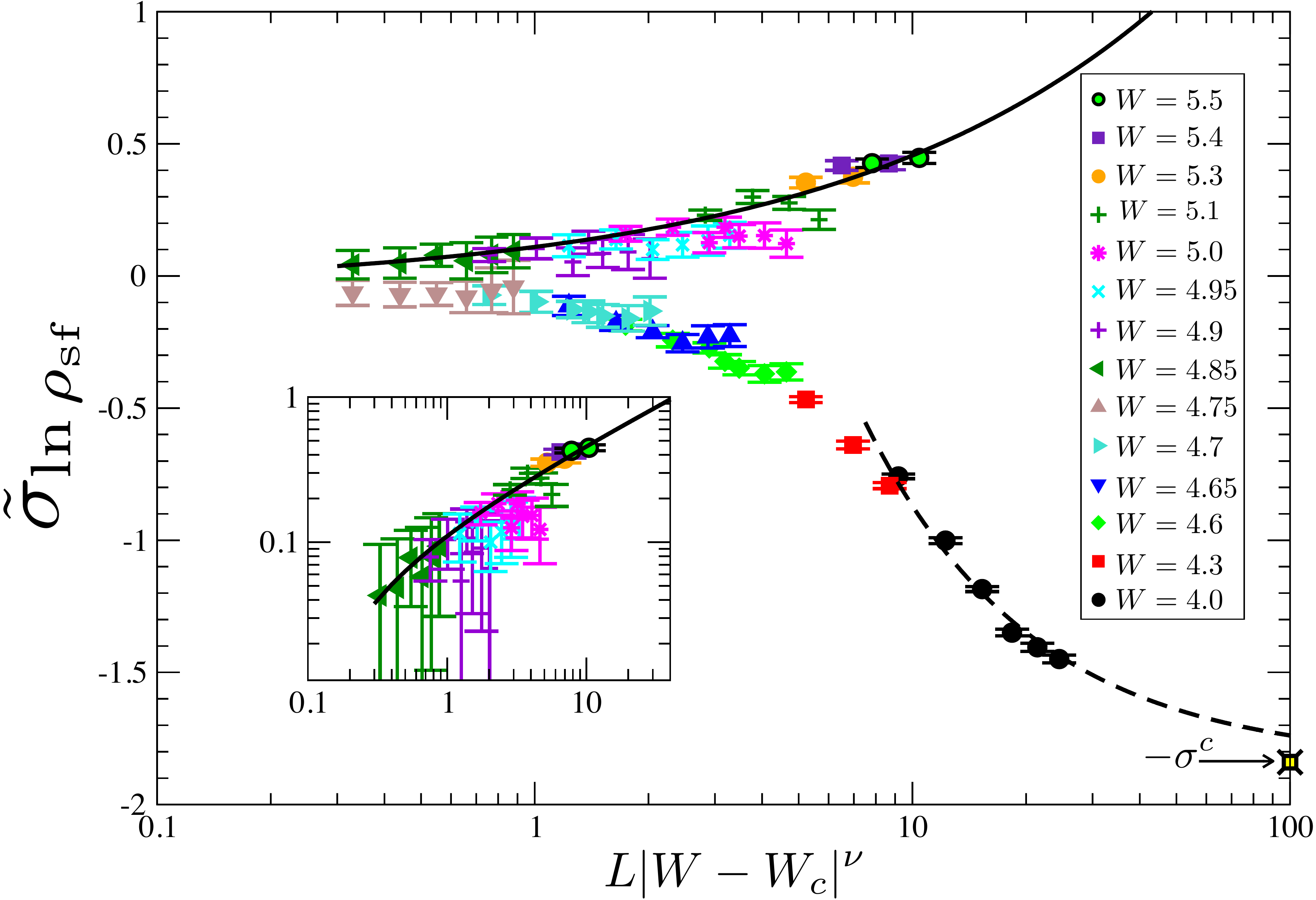}
       \caption{Corrected standard deviation of the logarithm of the SF response ${\tilde{\sigma}}_{\ln\rho_{\rm sf}}={{\sigma}}_{\ln\rho_{\rm sf}}-{{\sigma}}^{c}$ {\it{vs.}} system size in
           units of the typical length scale $\xi=|W-W_c|^{-\nu}$. In the SF phase $W< 4.8$, $\tilde\sigma$ tends to $-\sigma^{c}$ as $1/L$ (dashed line), whereas in the BG regime $W>4.8$, ${\tilde{\sigma}}$ grows as $L^{\omega}$ (full black line) with $\omega=0.5(2)$. Inset: zoom on the BG regime.}
\label{fig:5} \end{figure}

Coming back to the distributions, the drift observed for the typical stiffness in Fig.~\ref{fig:4}
(a) is related to the transient (irrelevant) broadening of $P(\ln\rho_{\rm sf})$ observed at small sizes in
Fig.~\ref{fig:3} (b). In order to take such a crossover into account and get rid of irrelevant corrections, we study the broadening of $P(\ln\rho_{\rm sf})$ using the corrected standard deviation (StD)
${\tilde{\sigma}}_{\ln\rho_{\rm sf}}={{\sigma}}_{\ln\rho_{\rm sf}}-\sigma^c$, where 
$\sigma^c$ is the StD at criticality.
This is plotted in
Fig.~\ref{fig:5} {\it{vs.}} $L|W-W_c|^\nu=L/\xi$, where a very good collapse
of the data is achieved without any adjusted parameters. 
In the SF regime, $\tilde\sigma$ converges towards $-\sigma^{c}$ as $1/\sqrt N$ (dashed curve), a consequence of self-averaging. More interestingly, the BG phase features an opposite qualitative behavior with $\tilde \sigma$
growing with system size, as $(L/\xi)^\omega$ (full line). A careful study of such very broad distributions hits the limits of our numerics, leading to quite large statistical errors, despite the very large number of samples ${\cal N}_s={\cal{O}}(10^4)$, but nevertheless allows to estimate the exponent $\omega=0.5(2)$. We interpret this result as follows: The prediction \cite{seibold_superfluid_2012} that the stiffness is dominated by quasi 1D paths suggests that one may understand the global SF response $\rho_{\rm sf}$ as a purely local quantity in the BG insulator. Moreover, an analogy~\cite{feigelman_superconductor-insulator_2010,lemarie_universal_2013} between the BG and the disordered phase of the random transverse-field Ising model~\cite{fisher_critical_1995}, as supported by recent 1D results~\cite{javan_mard_strong-disorder_2014}, suggests that the BG is governed by directed-polymer physics in dimension $1+1$~\cite{monthus2012random}. This predicts an exponent $\omega=1/3$~\cite{PhysRevLett.55.2924} for local quantities which is compatible with our estimate.

\begin{figure}[t] \centering \includegraphics[width=.95\columnwidth,clip]{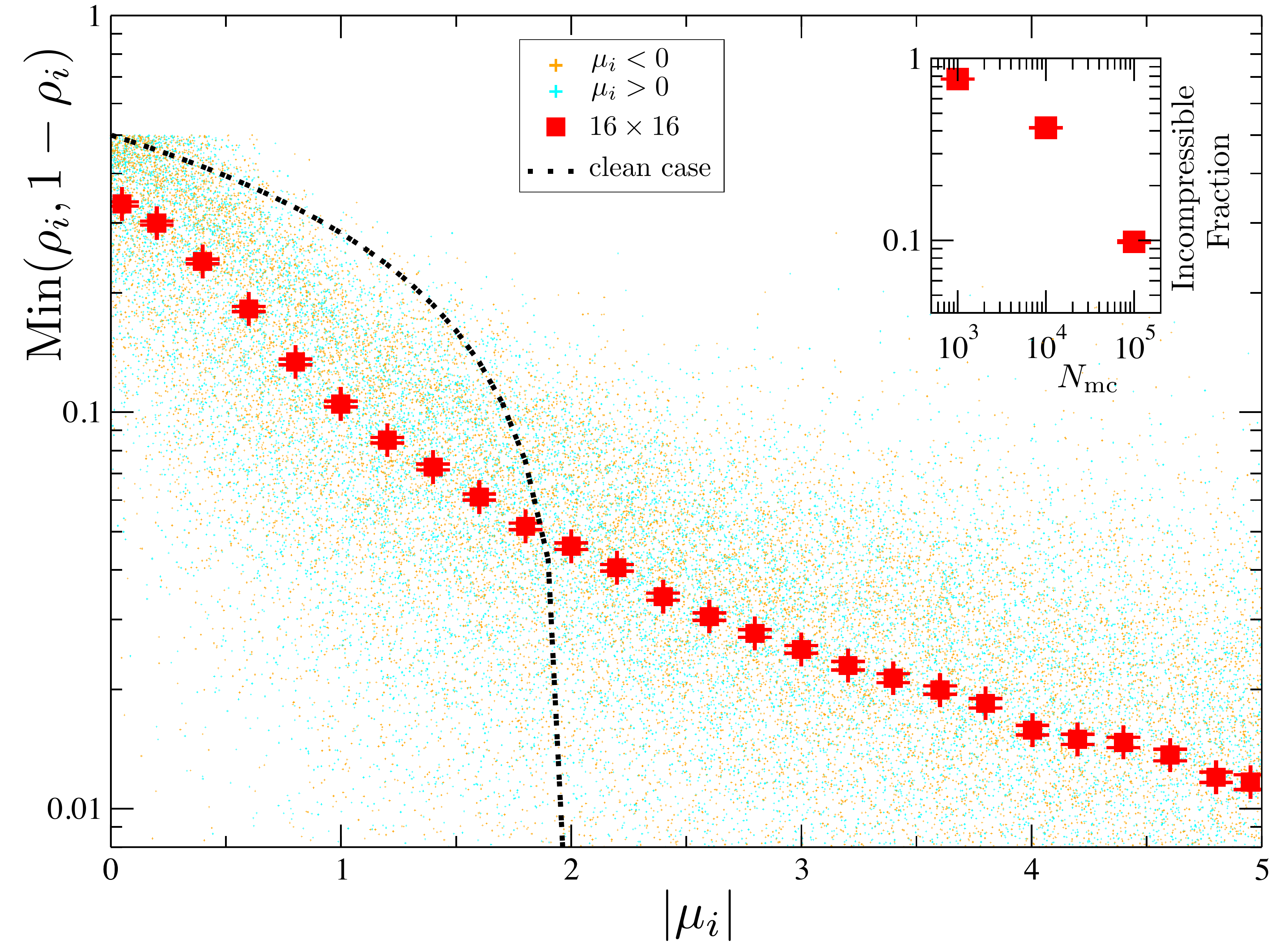}
       \caption{
           Local particle (hole) densities $\rho_i$ ($1-\rho_i$) plotted {\it{vs.}} local chemical potentials $|\mu_i|$. 
       Small blue (orange) points show QMC results of ${\cal{N}}_{\rm s}=150$ samples of size $16\times 16$ measured
       at $\beta t=1024$ with $N_{\rm mc}=10^5$ steps. 
       The red data points show averages over windows of the chemical potential of size
       ${\overline{\mu}}-0.1\le\mu_i\le{\overline{\mu}}+0.1$ and the clean ($W=0$) result is shown
       by the dashed line, yielding exactly zero for $|\mu|>2$ for $\text{Min}(\rho_i,1-\rho_i)$~\cite{coletta_semiclassical_2012}. 
        The inset quantifies the incompressible fraction, \textit{i.e.} the fraction of sites with
        $\rho_i=0$ or $\rho_i=1$ as a function of MC steps showing that in the
        exact limit of infinite Markov chains the incompressible fraction tends to zero.
        \label{fig:6} } 
    \end{figure}

\noindent {\it{Local density and absence of percolation---}} Finally, we want to discuss some microscopic properties of the insulating BG state. For this we focus on the local bosonic density $\rho_i=\langle b^{\dagger}_i b^{\vphantom\dagger}_i\rangle$, shown in Fig.~\ref{fig:6} in the BG regime ($W=5$) for $16\times 16$, at low enough temperature $\beta t=1024$ such that the total number of bosons does not fluctuate (see \cite{sm}). 
Clearly, the average behavior is
always compressible, which contrasts with the clean case where the system is incompressible whenever $|\mu|> 2$~\cite{coletta_semiclassical_2012}. Furthermore, the
fraction of incompressible sites with $\rho_i=0$ or $1$ decreases with the number of MC steps
and seems to vanish in the exact limit (inset). This shows that percolation through
compressible sites is present even in the BG phase, at least from such a single particle view, and is therefore not related to the SF-BG transition, in contrast with some recent discussions~\cite{niederle2013superfluid,krinner_direct_2013}.\\
\\
{\it{Conclusions---}} Large-scale QMC simulations of the SF-BG transition supplemented by finite-size scaling show that SF and BEC densities disappear at the same critical disorder strength $W_c\approx4.8$, with critical exponents $z\approx1.85$, $\nu\approx1.2$, and $\eta\approx -0.4$. The SF density distribution becomes infinitely broad upon increasing system size in the BG insulator, a characteristic signature of the absence of self-averaging supporting the fact that the SF density is a purely local quantity at strong disorder.  Our results also rule out a classical percolation scenario of incompressible sites in the BG. \\
\\
This work was performed using HPC
resources from GENCI (grant x2014050225) and CALMIP (grant 2014-P0677), and is supported by the
French ANR program ANR-11-IS04-005-01 and by the label NEXT. During the completion of this study we became aware of a parallel work~\cite{erik}, which reaches comparable conclusions.
\bibliographystyle{apsrev4-1}

\appendix
\clearpage
\section{SUPPLEMENTARY MATERIAL}

    This supplementary material provides additional information about our results and details the
    simulation and data analysis methods. We discuss in detail how finite temperature QMC results
    are used to access ground-state information, how we perform our careful error analysis and how we
    extract our best estimates for the critical exponents of the transition. We also analyze the
    problem of large autocorrelation times for the superfluid density in the Bose glass phase.

\begin{figure}[b]
    \centering
    \includegraphics[width=1\columnwidth]{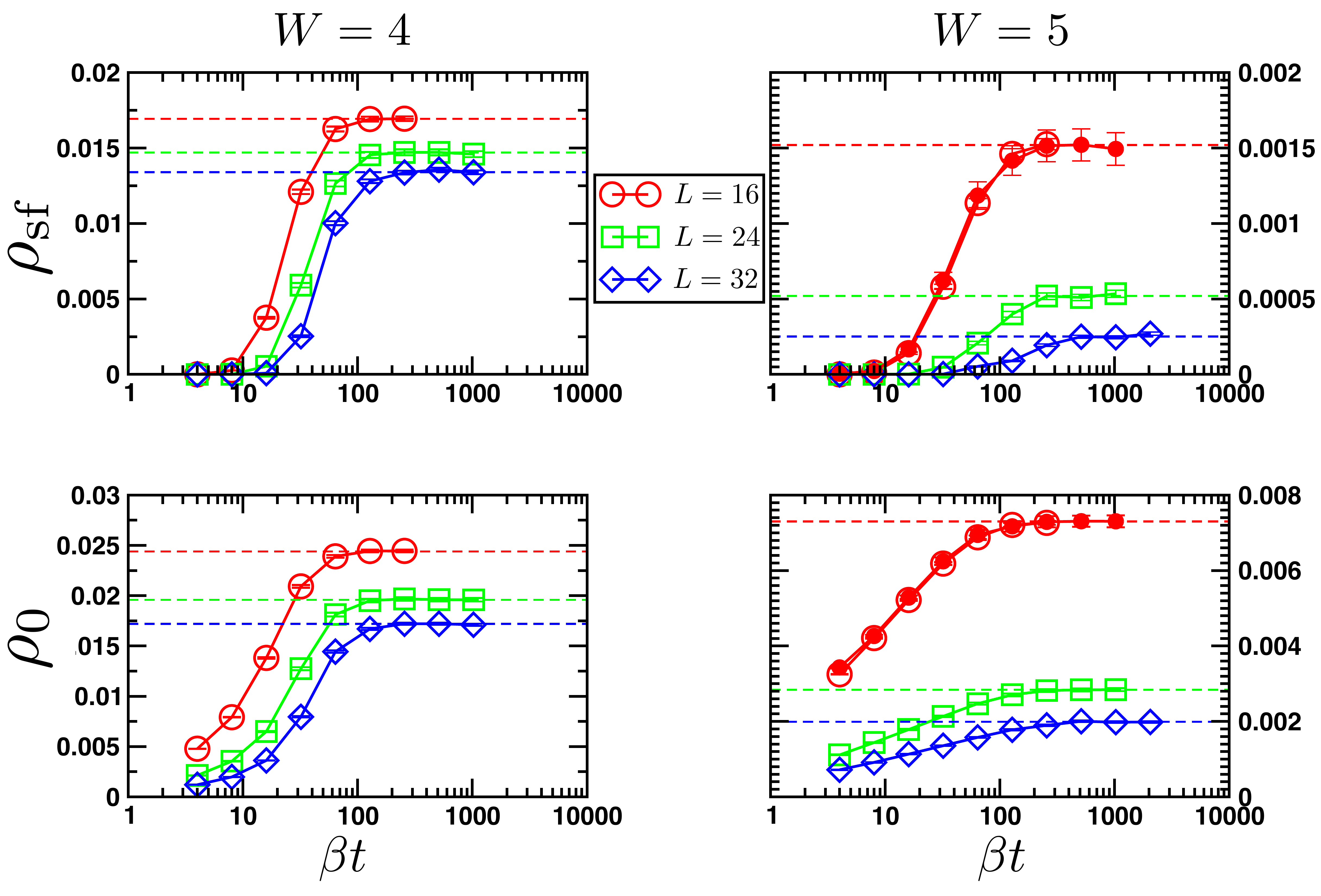}
    \caption{Disorder average of $\rho_{\text{sf}}$ (top) and $\rho_0$ (bottom) as a function of inverse temperature $\beta t$ for
different disorder strengths and system sizes. Open symbols are for simulations with
$N_{\text{mc}}=10^3$ Monte Carlo steps and filled symbols ($L=16$, $W=5$ ) for $N_{\text{mc}}=10^4$
Monte Carlo steps. The averages for different number of MC steps are in full agreement. The average densities saturate to their $T=0$ value (indicated by the dashed lines) at a finite $\beta t$ which will be used to perform the production ground-state simulations.}
    \label{fig:beta_dep}
\end{figure}

\section{I. $\beta$-DOUBLING SCHEME AND GROUND-STATE CONVERGENCE}
The stochastic series expansion (SSE) being a finite temperature method, it is
  important to perform calculations at low enough temperatures to capture
  ground state (GS) properties.
In order to accurately
  study this GS convergence, we have used the $\beta$-doubling scheme~\cite{sandvik_classical_2002_sm} which consists in performing simulations at exponentially decreasing temperatures by
  doubling the inverse temperature $\beta$ as long as we still see a change of
  the disorder average between the two previous temperatures. 
  
  This $\beta$-doubling trick
   allows us 
  to determine the inverse temperature $\beta t$ at which we should perform the simulations 
  for every system size and every disorder strength in order to ensure that we are really investigating
  the GS properties. In Fig.~\ref{fig:beta_dep} we show for two disorder strengths: $W=4$ in the superfluid (SF)  and $W=5$ in the Bose glass (BG) results of $\beta$-doublings for SF and BEC densities, both averaged over $\sim 10^3$ random samples for three system sizes ($L=16,~24,~32$). We nicely see that GS-converged expectation values are reached when saturation is achieved. From this $\beta$-doubling study, we have fixed for the production simulations 
  $\beta t=2^7$ for $L=12$ and up to $\beta t=2^9$ for $L=32$ to study the quantum ($T=0$) superfluid - Bose glass phase transition.

  It is worth noticing that in the right panel of 
  Fig.~\ref{fig:beta_dep} we also show the $\beta$-doubling data for $L=16$ and $W=5$ with 
  a bigger number of MC steps, $N_{\text{mc}}=10^4$, which are in perfect agreement with the
  results for $N_{\text{mc}}=10^3$. This feature will be further discussed below. 

  Note that the $\beta$-doubling scheme can be implemented efficiently in SSE by replicating
  (and inverting) the operator string at inverse temperature $\beta t$ to obtain
  an operator string twice as long as an efficient starting point at the inverse
  temperature $2\beta t$, thus minimizing the equilibration time of the Markov
  chain at $2\beta t$. This method is also used to speed up the thermalization
  process for all our production runs. 
 
\begin{figure*}[t]
    \centering
    \includegraphics[width=1.8\columnwidth]{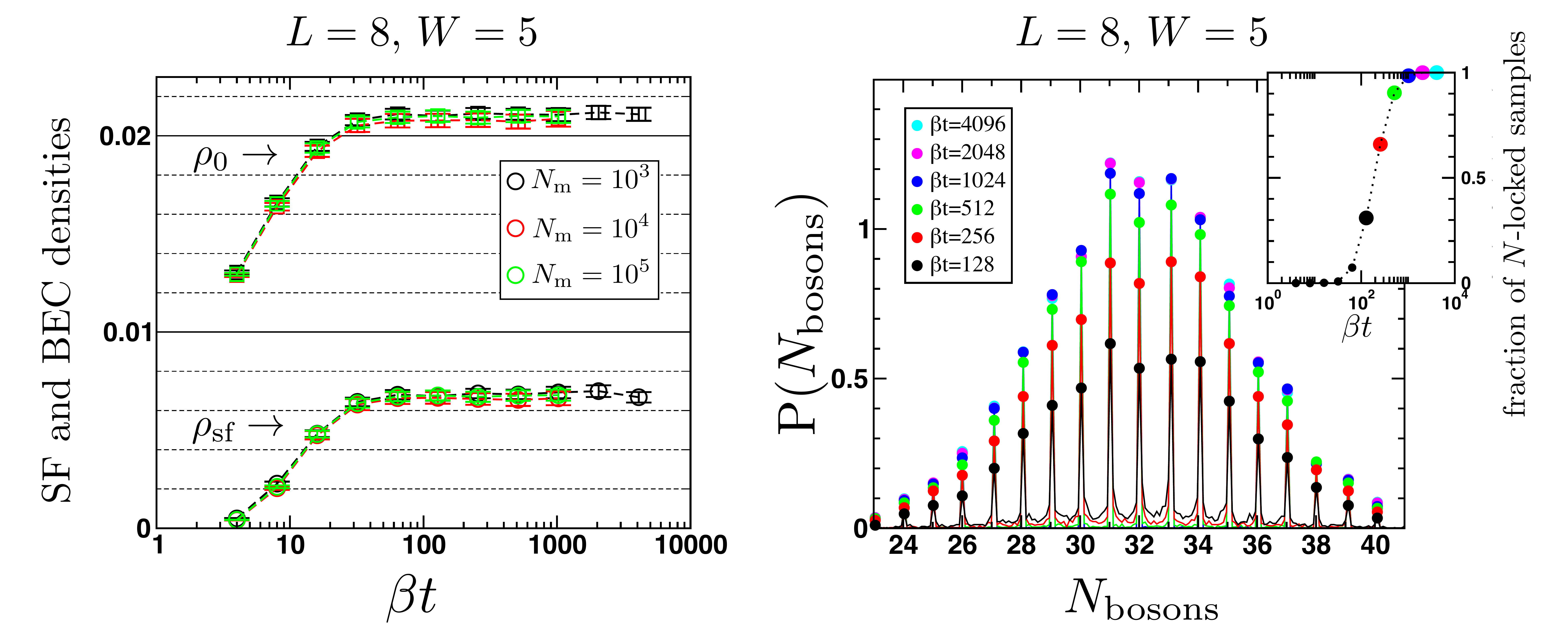}
    \caption{Left panel: Disorder averaged $\rho_{\text{sf}}$ and $\rho_0$ as a function of inverse
    temperature $\beta t$ for different number of Monte Carlo steps $N_{\text{MC}}$ for a system of
size $N=8\times8$ at disorder strength $W=5$, averaged over several hundreds of samples. The curves for different number of MC steps are in
full agreement. Right panel: particle number histogram for several hundreds of disorder
realizations and different temperatures for systems size $N=8\times8$. The inset shows the fraction
of disorder samples that are actually locked in a sector of fixed number of bosons as a function of
the inverse temperature.}
    \label{fig:density}
\end{figure*}
 
  Nevertheless, the notion of GS convergence appears to be quite subtle when monitoring various observables. Indeed, the temperature at which a given disordered sample
  is effectively in its GS, {\it{i.e.}} below the finite size gap,
  is strongly tied to the disorder realization, and contributions from low energy excited states 
  depend on the physical observable which is measured. For example, the total number of particles $N_{\rm bosons}=\langle\sum_i n_i\rangle$ is a good quantum number and therefore has to be locked in the GS to an integer number. In the Bose glass (BG) regime, where very small finite size gaps are expected,  the histogram of $N_{\rm bosons}$ obtained over several hundreds of samples slowly evolves when varying the temperature towards a collection of $\delta$-peaks, but only for very large $\beta$, as shown in Fig.~\ref{fig:density} (right panel).
Interestingly, the fraction of samples that are not fully converged to their GS (inset of
Fig.~\ref{fig:density} right), as far as the total number of bosons is concerned, remain sizeable while the disorder average superfluid (SF) or Bose-Einstein condensate (BEC) densities appear well converged to their $T=0$ values, as shown in the left panel of Fig.~\ref{fig:density}. There we see that for $N=8\times8$, the average SF and BEC densities appear converged in temperature for $\beta t \geq 128$ while the actual fraction of samples 
  with $N_{\rm bosons}$ locked is less than 30\% for $\beta t=128$.
  
  BEC and SF densities converge much faster to their $T=0$ values than $N_{\rm bosons}$. This undoubtedly facilitates the GS simulations as we can stop the $\beta$ cooling procedure at not too large inverse temperature, as far as SF and BEC densities are concerned. Nevertheless, one should still pay attention to potential systematic bias that may be introduced by rare disorder
realizations that may have not fully converged to their GS values for $\rho_{\rm sf}$ and $\rho_0$. To do so, we fit the $\beta$-doubling curves of $\rho_{\text{sf|0}}$ as a function of temperature to the following form 
\be
\rho_{\text{sf}|0}(\beta)=\rho_{\text{sf}|0}(\beta\to\infty)-A_{\text{sf}|0}\exp\left({-\beta/\beta_{\text{sf}|0}}\right),
\ee
which turns out to describe quite well our results with the following fitting parameters
\be
\left\{
\begin{aligned}
A_{\text{sf}}\approx&\frac{2}{L^2},\\
A_{0}\approx&\frac{1}{L^2},\\
\beta_{\text{sf}}=&\beta_{0}\approx0.1L^2\text{.}
\end{aligned}
\right.
\ee
With this simple phenomenological description, we have checked the stability of our scaling analysis, and concluded that our results do not
change upon the inclusion of such a correction term (see Fig. \ref{fig:bs_estimates}). Note that the
correction is small enough so that the corrected result is still within the error bar as guaranteed
by our convergence check in Fig. \ref{fig:beta_dep}.

\section{II. ERROR BARS AND EVALUATION OF SYSTEMATIC ERRORS}
  \begin{figure*}[t]
    \centering
        \includegraphics[width=1.3\columnwidth]{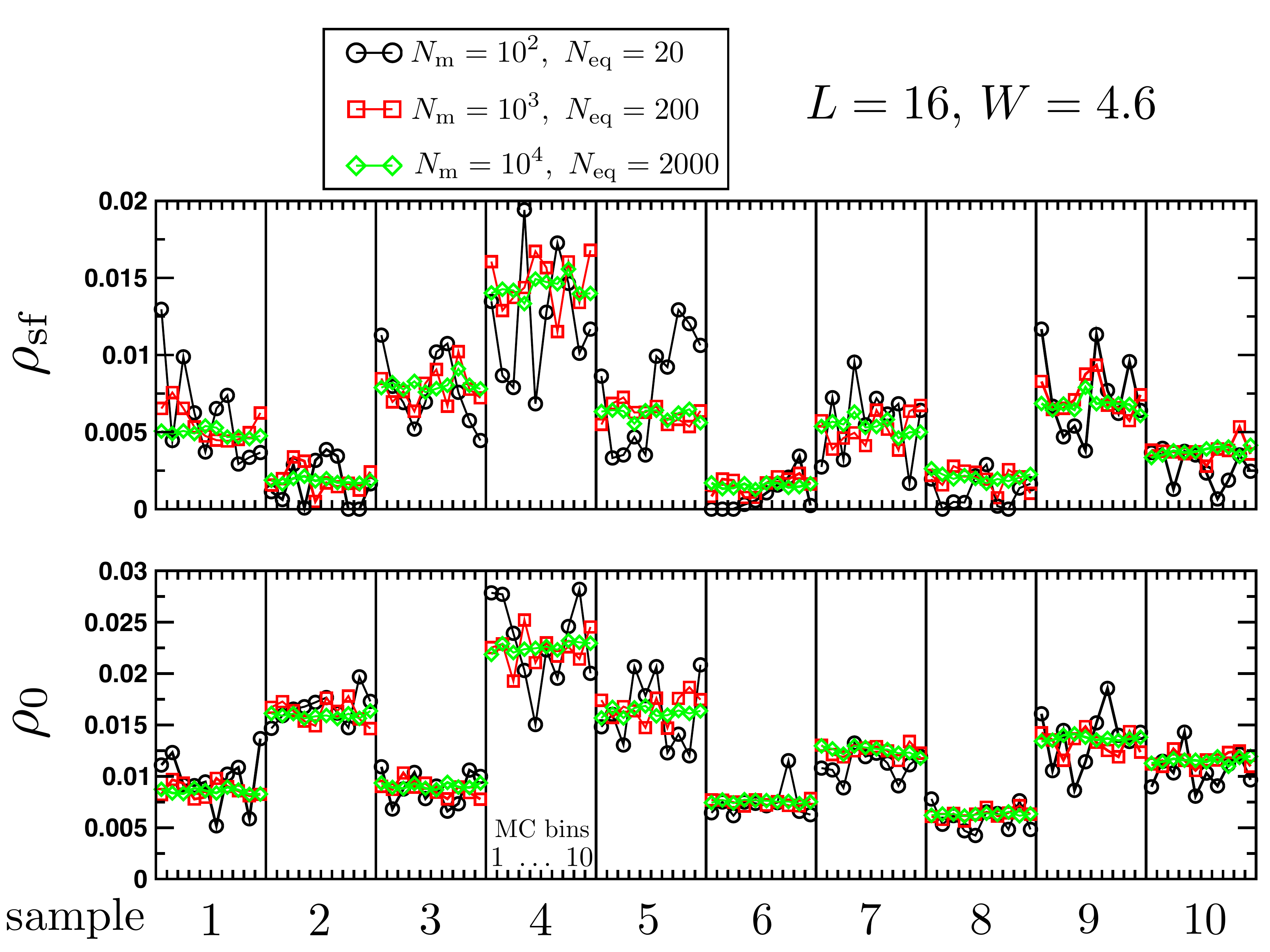}
    \caption{ Example of 10 disorder realizations for a system of size $N=16\times16$ at disorder
        strength $W=4.6$ (SF regime). The upper (lower) panel shows the SF (BEC) density measured for each disordered
        sample with 10 independent consecutive bins, each with a different number of MC Steps $N_{\text{m}}=10^2, \, 10^3,\,10^4$ (different symbols).
        It is clear that for $N_{\text{m}}\geq 10^3$ the MC fluctuations are smaller than
        fluctuations between disordered samples.}
    \label{fig:mc_vs_dis}
\end{figure*}

  Monte Carlo (MC) results for disordered systems have two sources of
  statistical errors: (i) Statistical fluctuations of the Monte Carlo result for
  each random sample, which can be reduced by generating longer Markov
  chains. (ii) Statistical fluctuations of the disorder average over random
 samples which can be reduced by including more realizations.

\subsection{A. MC error vs. disorder fluctuations}

  It is crucial to distribute the available computer time efficiently over the
  different competing tasks (large $\beta$, number of MC steps, number of random samples) in order to maximize the overall precision.
  Therefore, we compare the MC error bar to the error bar stemming from
  disorder averaging. We choose a minimal number of 1000 MC measurements in order to assure that the
Markov chain is much longer than the autocorrelation times of $\rho_0$ and
$\rho_{\mathrm{sf}}$ and it is not allowed to perform less MC steps as this
leads to wrong results and introduces a systematic error.
  
  \begin{figure}[b]
    \centering
        \includegraphics[width=\columnwidth]{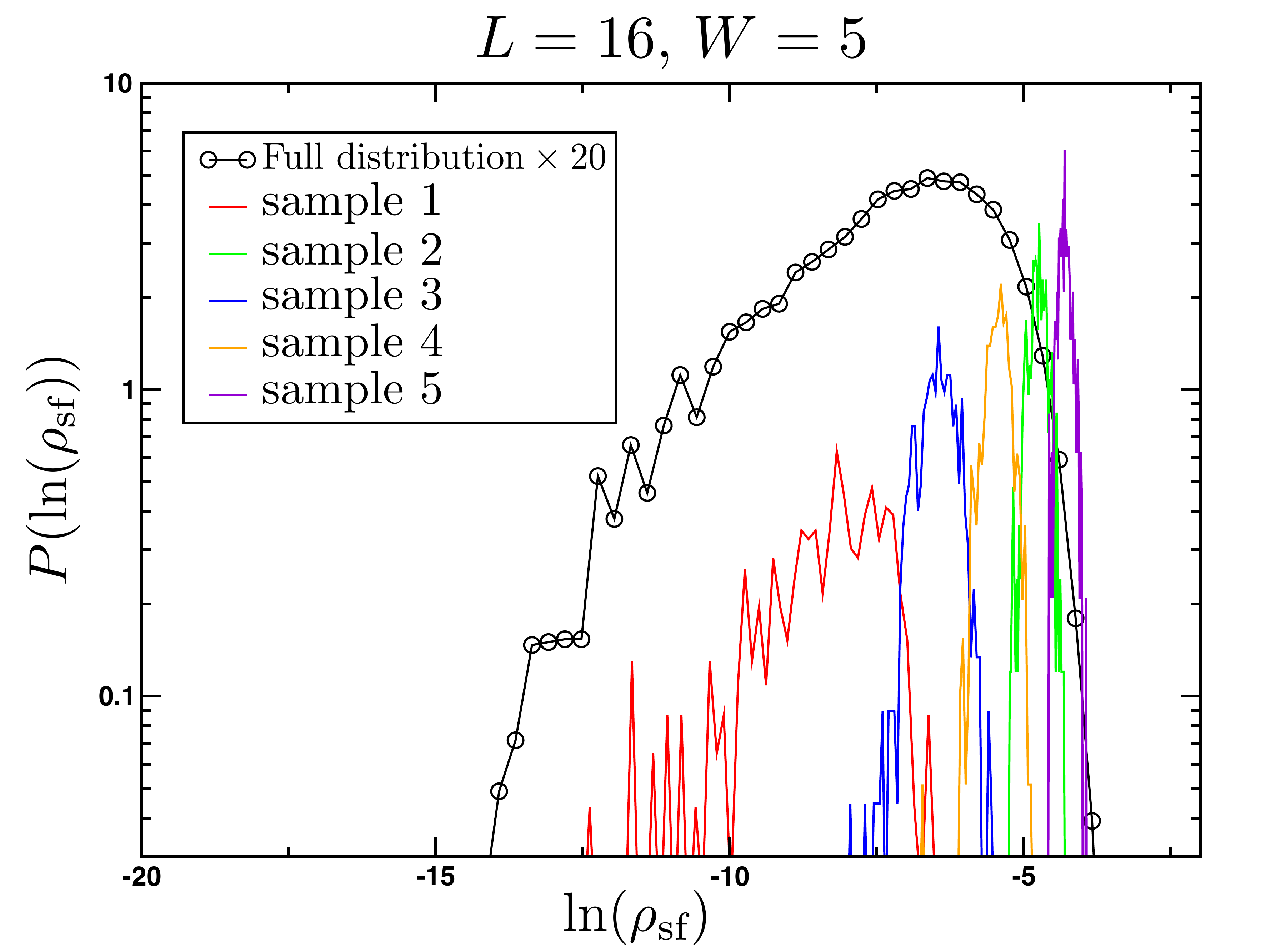}
    \caption{Distribution of $\ln(\rho_{\text{sf}})$ for system size $N=16\times16$ at disorder strength $W=5$ (BG). The black circles curve 
    is the full distribution over $\approx 20000$ disordered samples, multiplied by a factor of 20
    for graphical reasons. The colored lines are the histograms of $\sim 300-400$  MC bins of
    $N_{\rm m}=10^3$ steps, shown for 5 representative disordered samples. The spread of the MC bin distributions is clearly
    much smaller than the spread stemming from disordered samples.  }
    \label{fig:mc_vs_dis_hist}
\end{figure}

  Figure~\ref{fig:mc_vs_dis} shows clearly that MC
  fluctuations within one disorder realization are much smaller than the
  error bar stemming from the fluctuations between disorder realizations. Indeed, we see that while MC fluctuations with $N_m=100$ are of the same order of magnitude that sample-to-sample fluctuations, making $N_m=1000$ measurement steps is enough to keep MC errors much smaller than fluctuations due to random configurations.To further illustrate in a more quantitative way the fact that disorder fluctuations are much bigger 
than MC fluctuations, we show in Fig.~\ref{fig:mc_vs_dis_hist} the full distribution of $\ln(\rho_{\text{sf}})$ obtained for 
a system of linear size $L=16$ with ${\cal N}_s\approx 20000$ samples at disorder strength $W=5$ (BG
regime) on which we have superimposed histograms of $\sim 300-400$ MC averages over $10^3$ MC steps for 5 representative samples.
Hence, it appears more efficient to perform a relatively modest number of MC steps $\sim 10^3$, in order to be able to
  sample more disorder realizations. Note that the MC error is nevertheless
  included in our data analysis as discussed below.

\subsection{B. Disorder fluctuations}
The problem of increasing variances in the BG phase leads to a sampling issue which can be tackled
for finite systems by using a very large number of disorder realizations. In order to estimate how many
samples should be used to obtain a reliable and converged result, we calculate running means as a
function of the number of disorder realizations.

\begin{figure*}[t]
    \centering
        \includegraphics[width=1.4\columnwidth]{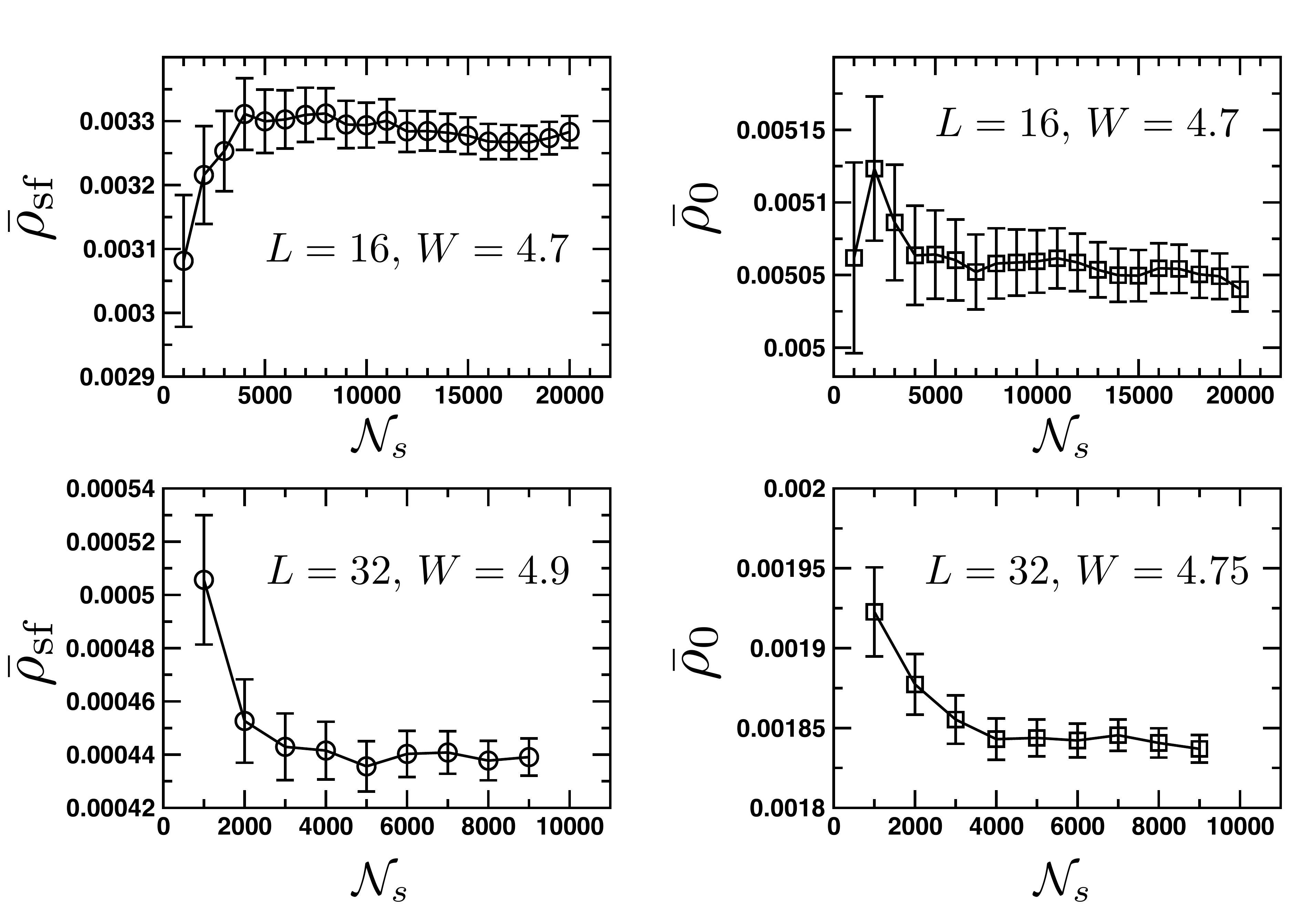}
    \caption{Disorder average of the superfluid density $\rho_{\mathrm{sf}}$ and the Bose condensed density  $\rho_0$  as a function of number of disorder realizations for different system sizes ($N=16\times16$, $32\times32$) and disorder strengths. A number of samples ${\cal{N}}_s \gg 10^3$ appears necessary to avoid unconverged disorder averages.}
    \label{fig:disavg}
\end{figure*}

Figure~\ref{fig:disavg} shows the estimated averages and error bars of both SF
and BEC densities for increasing number of disorder realizations for different system sizes and disorder strengths.
Seemingly, averages over around 1000 disorder realizations are not fully converged and 
for the smaller system sizes, at least some 10000 disordered samples are needed
for a converged result, while for larger systems, at least 5000 disordered samples are required. To
be safe, we choose to reach $\sim$20000 disorder realizations for the smaller sizes $L\leq22$ and $\sim$10000 for system sizes $L\geq 24$.      

In addition, a very large number of disorder realizations are also needed to correctly sample the distributions of $\ln(\rho_{\text{sf}})$
since they broaden with increasing system size, as shown in the main text. This further justifies
our choice of such very large numbers of samples 
(see also Ref.~\cite{lin_superfluid-insulator_2011_sm} for a related discussion).

\subsection{C. Bootstrap analysis}

For our scaling analysis of $\rho_0$ and $\rho_{\text{sf}}$ it is crucial to
take into account the correct error bars of our results for every system size and
disorder strength. While the average value is simply given by the disorder
average of the MC averages, the total error bar can be estimated using a
bootstrap approach. For this, we generate a set of bootstrap samples by randomly
selecting a subset $\{ i_B \}$ from the $\mathcal{N}_s$ realizations
with replacement (selecting in total $\mathcal{N}_s$ realizations from the ensemble with replacement)
and drawing a Gaussian random number distributed according to 
\begin{equation}
    p(x) = \frac{1}{\sqrt{2\pi}\sigma_\text{MC}} \text{e}^{ -(x-\rho)^2/(2 \sigma_\text{MC}^2) }
\end{equation}
for each selected sample, which we then average over $\{ i_B \}$. This is
repeated many (typically $\approx1000$) times and the standard deviation of the
result is indeed an accurate estimator of the total error bar. We have also
checked that the MC error is smaller than the disorder
fluctuations so that the final results are unchanged if it is neglected (see Fig. \ref{fig:bs_estimates}).

In order to determine error bars of the critical disorder strength $W_c$ and the
critical exponents, a second level of bootstrap analysis is introduced, which
performs multiple fits by a gaussian resampling of our results for $\rho_0$ and
$\rho_\text{sf}$ within the previously determined error bars. The standard
deviation of the such obtained fit results represent the statistical error of
the final results (see Fig. \ref{fig:bs_estimates} and table \ref{tab:S1})

\section{III. OVERCOMING LARGE AUTOCORRELATION TIMES FOR $\rho_{\text{sf}}$}

\subsection{A. Dynamical increase of the number of MC steps}
\begin{figure*}[t]
    \centering
    \includegraphics[width=1.5\columnwidth]{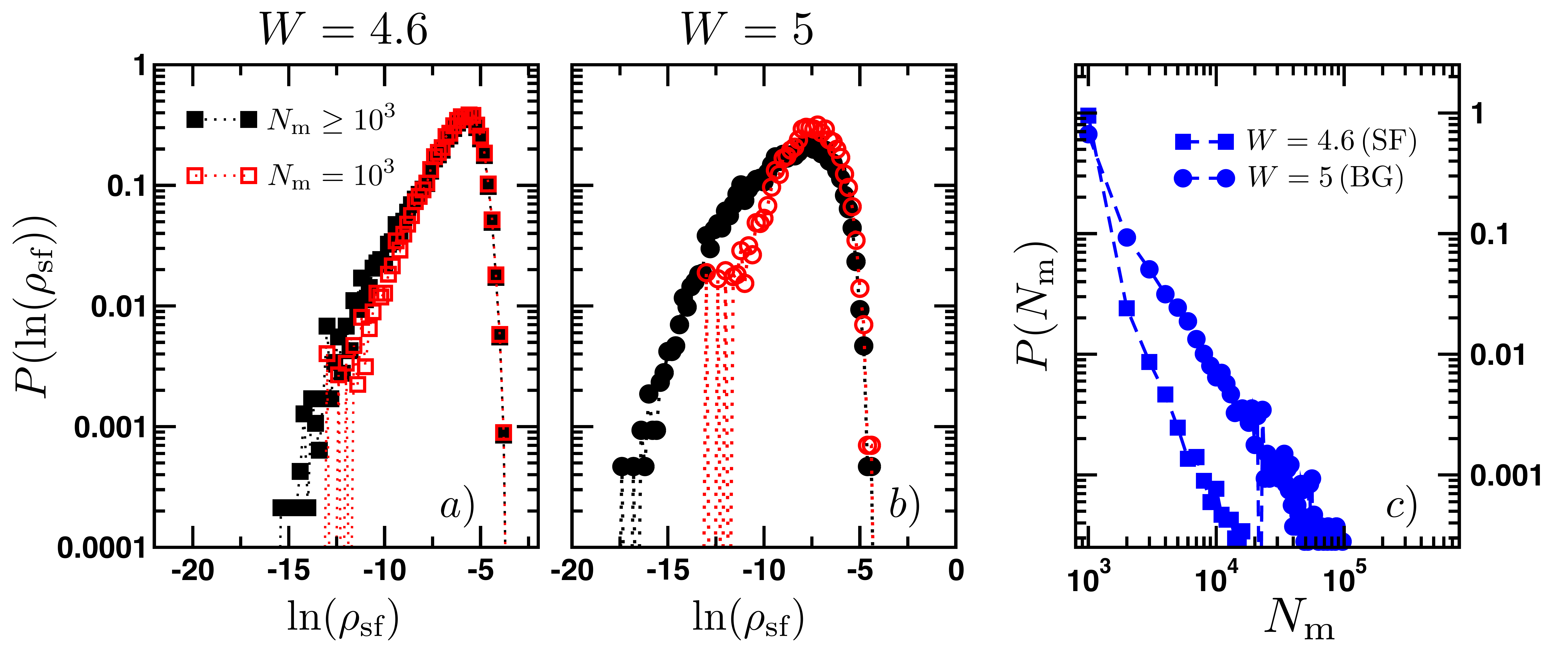}
    \caption{Histogram of $\ln(\rho_{\text{sf}})$ in SF (panel $a$) and BG (panel $b$) phase 
    for a system of size $N=24\times24$ when the number of MC steps is fixed to $10^3$ (red) 
    and when it is dynamically adjusted as described in the text (black). c) Fraction of disorder 
    realizations having needed $N_{\text{mc}}$ Monte Carlo steps for the same system size and disorder strengths. }
    \label{fig:bins_histo}
\end{figure*}

For disorder realizations which exhibit a particularly small value of the
superfluid density $\rho_{\text{sf}}$, which is measured in the SSE by counting
the winding number fluctuation of worldlines, we find that autocorrelation times become
larger and the change between the $0$ winding number sector to nonzero
winding numbers takes more MC steps. For these cases, we decide dynamically to
perform additional blocks of 1000 MC steps until the final result is reliable
and the total simulation time is much longer than the autocorrelation time,
involving simulations with up to $10^5$ MC steps. This method gives reliable access to small values of $\rho_{\text{sf}}$, which
would be estimated to be $0$ otherwise, which is not expected for our finite size samples.

We set the $10^5$ MC steps limit 
for computing time reasons and the rare disorder realizations needing more steps
are evaluated to zero stiffness. These realizations cannot be used for the discussion 
on the distributions of $\ln(\rho_{\text{sf}})$, nor the typical stiffness. However, they can be included
in the calculation of the average stiffness as neglecting them would induce systematic errors. In
fact, our analysis leads us to speculate that these samples have been simulated at a too high 
temperature and are not converged to the GS.

There are several possibilities which we all explored and found little dependence of our results on
the particular choice. 
\textit{i)} Adding samples of vanishing $\rho_{\text{sf}}$ to the data set and performing the data analysis 
described before. This corresponds to approximating the distribution of $\rho_{\text{sf}}$ by a
distribution with a cut-off at the minimal stiffness observed in the ensemble and a delta peak at
zero. Clearly this is an incorrect estimation as the correct distribution vanishes at zero and the
delta peak is both due to insufficient simulation time and too high temperature, thus introducing a
(small) systematic error.
 \textit{ii)} 
Extrapolation of the distributions of $\ln(\rho_{\text{sf}})$ towards zero by a power law tail (straight line in the plots in figure~\ref{fig:bins_histo}) 
but since this is the least properly sampled part of the distribution a reliable extrapolation cannot be achieved. 
\textit{iii)} The replacement of the non physical zero values by half the value of the cut-off (\textit{i.e.} the minimal computed stiffness for every
 couple $(L,W)$).  This corresponds to approximating
  the small $\rho_{\text{sf}}$ tail of the distribution by a box distribution and is only justified by
  the very small weight of the approximated part of the distribution.
We used this last solution for checking the stability of our scaling analysis and found that our results
  are not dependent on whether these samples are included or not.
Hence, to treat on equal footing the typical and average stiffness, we decided
to neglect them, as the introduced bias is much smaller than our overall uncertainty. \\
\\

Future work
could try to adress this problem by investing more computer time in samples with small values of
$\rho_{\text{sf}}$. 
We have found that these samples are typically not completely converged in
the GS and it should be explored if one can solve the problem by reducing the temperature
iteratively in such samples.

\subsection{B. Improved sampling of the full stiffness distributions}
A comparison of results obtained
with only $N_m$ fixed to $1000$ MC steps to results obtained with our dynamic Markov chain
length method are displayed in Fig.~\ref{fig:bins_histo} for a system in the SF (though
strongly disordered) regime (panel $a$) and inside the BG phase (panel $b$). It is clear that our 
dynamical adjustment of the MC steps allows for much smaller SF stiffnesses to be computed,
especially inside the BG phase. The implementation of this procedure allows for the
 distribution of $\ln(\rho_{\text{sf}})$ to be accurately sampled and  hence, for an appropriate
 estimation of the standard deviation of $\ln(\rho_{\text{sf}})$. Without this method, the broadening 
 of the distributions for growing system sizes inside the BG phase (described in the main text)
 could not have been investigated quantitatively, nor even qualitatively observed.\\
 \\

\section{IV. CRITICAL EXPONENTS}
\begin{figure*}[t]
    \centering
 	  \includegraphics[width=1.3\columnwidth]{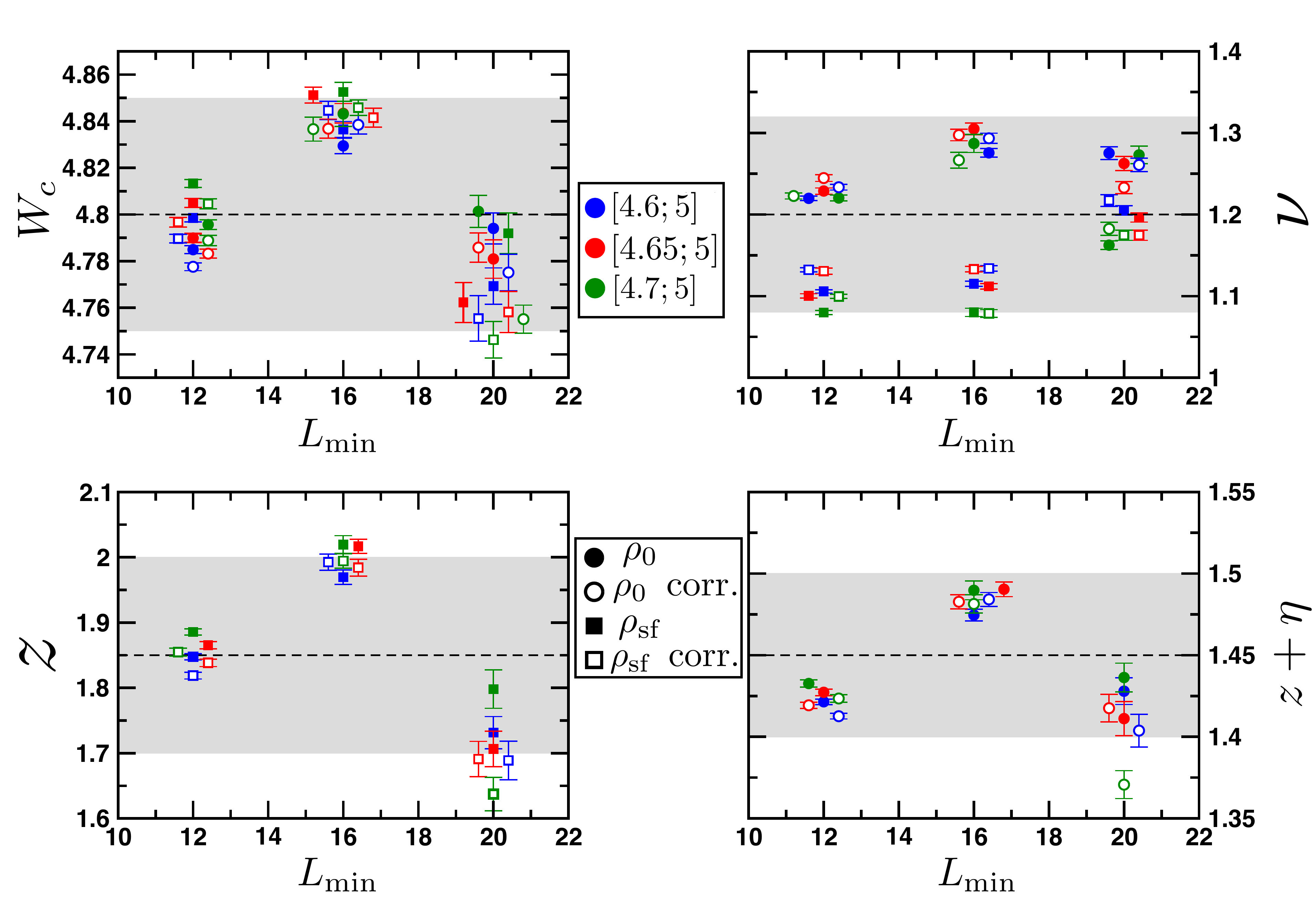}
    \caption{Bootstrap estimates of the critical exponents for different disorder and system size windows. Different colors represent the estimates for different windows of disorder. Circles depict estimates obtained with the scaling of the BEC density $\rho_0$ and squares those obtained via the superfluid density $\rho_{\text{sf}}$. Full symbols correspond to critical parameter estimates when no systematic error correction is added and open symbols estimates where both the Monte Carlo and temperature systematic error corrections are included. The dashed line gives the final best estimate and the shaded area the final uncertainty.}
    \label{fig:bs_estimates}
\end{figure*}
%
\begin{table}[h]
    \begin{tabular}{ l  |l   c  c | c  r  r  r   }
    \hline
    \hline
    &  \multicolumn{3}{c}{$Q_0$ ($\times 100$)} & \multicolumn{3}{|c}{$Q_{\mathrm{sf}}$ ($\times 100$)}  \\  
    \hline
$L_{\mathrm{min}}$ & $[4.6;5]$ & $[4.65;5]$ & $[4.7;5]$ & $[4.6;5]$ & $[4.65;5]$ & $[4.7;5]$\\ 
\hline
$12$ & 8.7 & 13.3 & 16.2 & 4.5 & 4.25 & 11\\ 
$16$ & 7 & 15.2 & 13.5 & 4 & 5 & 9.2\\   
$20$ & 21.3 & 39.75 & 33.8 & 37.5 & 39.25 & 39.5\\
\hline
    \hline
\end{tabular}
\caption{\label{tab:S1}Quality of the fits of the critical parameters from the Bose condensed density ($Q_0$) and from the superfluid density ($Q_{sf}$), corresponding to the parameters shown in table 1 in the main text and figure \ref{fig:bs_estimates} for different windows of disorder and of system size $[L_{\mathrm{min}};32]$.}
\end{table}

  We have discussed above how the statistical error on the critical disorder
  strength and the critical exponents may be obtained. However, we believe that
  the systematic error due to small system sizes which may require corrections to scaling in our
  scaling analysis is in fact dominant, as results
  change slightly if the size of the fit window in system size or included
  disorder strengths is varied.

  Therefore, we have performed a systematic analysis of this systematic error
  and give error bars that represent the total fluctuation of our results. This is represented in Fig. \ref{fig:bs_estimates}.

  In order to quantify the quality of our fits, we calculate the sum of squared residuals 
  \be
  \chi^2 = \sum_i \left(  (\rho_{\mathrm{sf}|0}^i - \mathrm{fit}(W_i,L) )/\sigma_i \right)^2
  \ee
  and obtain the probability $Q$ of finding a $\chi^2$ greater or equal than this value given the
  fit by
  \be
  Q=\frac{1}{\Gamma(n_\text{dof}/2)} \int_{\chi^2/2}^{\infty} \text{d}y y^{n_\text{dof}/2-1}
  \text{e}^{-y},
  \ee
  as explained in Ref. \onlinecite{young_everything_2012}.

  The corresponding qualities of fit $Q$ for each window in size and included disorder strengths are shown
  in table \ref{tab:S1}. There are two reasons why the qualities of fit are systematically larger for the window $[20;32]$. 
  First, reducing the number of system sizes included in the analysis while keeping the bigger sizes means 
  that there is less size dynamics, hence the fact that no drift term is included in the fits becomes more justified. 
  Secondly, the bigger system sizes have slightly bigger relative error bars so the corresponding $\chi^2$ 
  is smaller and the quality of fit $Q$ larger as these two quantities are very sensible to the error bars.
  Consequently, we treat all fit windows on equal footing to estimate averages and uncertainties and
  do not give a bigger weight to the estimates obtained with $L_{\text{min}}=20$.   

 In figure~\ref{fig:bs_estimates} we plot the critical parameters with their error bars, obtained from a bootstrap analysis with 100 
 bootstrap samples, for every window in system size and included disorder strength corresponding to table~\ref{tab:S1}.
 For comparison, the critical exponents and critical disorder strength estimates, with their error bars, given upon including
 the corrections to the systematic errors due to temperature convergence and the Monte Carlo error bars are also shown (open symbols).
 It is clear that including the correction of the possible systematic errors does not change the final 
  estimations of the critical parameters as well as their error bars, which are quite large.

\end{document}